\documentclass[showkeys]{revtex4}
                                                                                
\usepackage{algorithm}
\usepackage{algorithmic}
\usepackage[mathscr]{eucal}
\usepackage{hyperref}
\usepackage{amssymb}
\usepackage{amsfonts}
\usepackage{amsmath}
\usepackage{latexsym}
\usepackage{bm}
\usepackage{xspace}

\newcommand{\D}{{\cal D}}
\newcommand{\Part}{\mathscr{P}}
\newcommand{\RP}{R_{\mathscr{P}}}
\newcommand{\R}{{\mathscr{R}}}
\newcommand{\QP}{Q_{\mathscr{P}}}
\newcommand{\Q}{{\mathscr{Q}}}

\newcommand{\eps}{\epsilon}

\newcommand{\qed}{\hfill\rule{7pt}{7pt}}
\newtheorem{theorem}{Theorem}[section]
\newtheorem{lemma}[theorem]{Lemma}
\newtheorem{observation}[theorem]{Observation}
\newtheorem{conjecture}[theorem]{Conjecture}
\newtheorem{question}[theorem]{Question}
\newtheorem{corollary}[theorem]{Corollary}
\newenvironment{proof}{\noindent{\bf Proof:}}{\qed\medskip}

\newtheorem{definition}[theorem]{Definition}
\newtheorem{fact}[theorem]{Fact}

\begin{document}

\title{Improved Bounds on Quantum Learning Algorithms}
\date{November 18, 2004}

\author{Alp At\i c{\i}}
\email{atici@math.columbia.edu}
\affiliation{
Department of Mathematics\\ 
Columbia University\\
New York, NY 10027}

\author{Rocco~A. Servedio}
\email{rocco@cs.columbia.edu}
\affiliation{
Department of Computer Science\\
Columbia University\\
New York, NY 10027}

\begin{abstract}
In this article we give several new results on the complexity of algorithms that learn 
Boolean functions from quantum queries and quantum examples.
\smallskip

\textbullet\quad Hunziker {\em et al.}~\cite{HMPPR} conjectured that for any class $C$
of Boolean functions, the number of quantum black-box queries which are
required to exactly identify an unknown function from $C$ is 
$O(\frac{\log |C|}{\sqrt{{\hat{\gamma}}^{C}}})$, where
$\hat{\gamma}^{C}$ is a combinatorial parameter of the class $C.$
We essentially resolve this conjecture in the affirmative by giving 
a quantum algorithm that, for any class $C$, identifies any unknown function from
$C$ using $O(\frac{\log |C| \log \log |C|}{\sqrt{{\hat{\gamma}}^{C}}})$ quantum
black-box queries.
\smallskip

\textbullet\quad We consider a range of natural problems intermediate between the
exact learning problem (in which the learner must obtain all bits of
information about the black-box function) and the usual problem 
of computing a predicate (in which the learner must obtain 
only one bit of information about the black-box function).  
We give positive and negative results on 
when the quantum and classical query complexities of these intermediate
problems are polynomially related to each other.
\smallskip

\textbullet\quad Finally, we improve the known lower bounds on the number of quantum examples
(as opposed to quantum black-box queries)
required for $(\eps,\delta)$-PAC learning any concept class of 
Vapnik-Chervonenkis dimension $d$ over the domain $\{0,1\}^n$
from $\Omega({\frac d n})$ to $\Omega(\frac{1}{\epsilon}\log
\frac{1}{\delta}+d+\frac{\sqrt{d}}{\epsilon})$.
This new lower bound comes closer to matching known upper bounds for
classical PAC learning.
\end{abstract}

\pacs{03.67.Lx, 89.80.+h, 02.70.-c}
\keywords{quantum query algorithms, quantum computation, computational learning theory, PAC learning}
\maketitle

\section{Introduction}

\subsection{Motivation and Background}

A major focus of study in quantum computation is the power of quantum
algorithms to extract information from a ``black-box'' oracle for an unknown
Boolean function.  Many of the most powerful ideas for both algorithmic
results and lower bounds in quantum computing have emerged from this framework,
which has been studied for more than a decade.

The most frequently considered problem in this setting is to determine
whether or not the black-box oracle (which is typically
assumed to belong to some particular {\em a priori} fixed class $C$ of possible functions) 
has some specific property, such as being identically 0 \cite{BBBV,GROVER},
being exactly balanced between outputs 0 and 1 \cite{DJ}, or being invariant
under an XOR mask \cite{SIMON}.
However, as described below 
researchers have also studied several other problems in which the goal is to obtain
more than just one bit of information about the target black-box function:

\begin{description}
\item[Quantum exact learning from membership queries:] 
Servedio
and Gortler \cite{RSSG} initiated a systematic study of the quantum black-box
query complexity required to 
{\em exactly learn} any unknown function $c$ from a class $C$ of Boolean functions.
This is a natural quantum analogue of the standard classical model 
of {\em exact learning from membership queries} which was introduced 
in computational learning theory by Angluin \cite{ANGLUIN}.
This quantum exact learning model was also studied by Hunziker 
{\em et al.}~\cite{HMPPR} and by Ambainis {\em et al.}~\cite{AKMPY}, who 
gave a general upper bound on the quantum query complexity of learning any class $C$.

\item[PAC learning from quantum examples:]
In another line of related research, Bshouty and Jackson \cite{BSHJA} introduced a natural 
quantum analogue of Valiant's well-known Probably Approximately Correct (PAC) model of Boolean
function learning \cite{VALIANT} which is widely studied in computational learning theory.
\cite{RSSG} subsequently gave a $\Omega(d/n)$ lower bound on the number of quantum examples required
for any PAC learning algorithm for any class $C$ of Boolean functions
over $\{0,1\}^n$ which has Vapnik-Chervonenkis dimension $d.$
\end{description}

\subsection{Our Results}

In this paper we study three natural problems of quantum learning:  
(i) exact learning from quantum membership queries as described above;
(ii) learning a {\em partition} of a class of functions from quantum membership
queries (this is an intermediate problem
between the quantum exact learning problem and the well-studied problem of obtaining a single bit
of information about the target function), and (iii) quantum PAC learning 
as described above.
For each of these problems we give new bounds on the number of
quantum queries or examples 
that are required for learning.

For the quantum exact learning model, 
Hunziker {\em et al.}~\cite{HMPPR} conjectured that for any class $C$
of Boolean functions, the number of quantum black-box queries that are
required to exactly learn an unknown function from $C$ is 
$O(\frac{\log |C|}{\sqrt{{\hat{\gamma}}^{C}}})$, where
$\hat{\gamma}^{C}$ (defined
in Section~\ref{sec:previous}) 
is a combinatorial parameter of the class $C$.
We give a new quantum exact learning algorithm
based on a multi-target Grover search on a prescribed subset of the
inputs, and show that the query complexity for this algorithm is $O(\frac{\log
|C| \log \log |C|}{\sqrt{{\hat{\gamma}}^{C}}})$; this resolves the conjecture of
Hunziker {\em et al.}~\cite{HMPPR} up to a $\log \log |C|$ factor.  Our new bound
is incomparable with the upper bound of Ambainis {\em et al.}~\cite{AKMPY}, but as we show it
improves on this bound for a wide range of parameter settings.  
We also show that for every
class $C$ of Boolean functions, the query complexity of our generic algorithm is guaranteed to be
at most a (roughly) quadratic factor worse than the query complexity
of the {\em best} quantum algorithm for learning $C$ (which may be
tailored for the specific class $C$).

For our second problem, we study a more general problem which is intermediate
between learning the black-box function exactly and 
computing a single Boolean predicate of the unknown black-box function.
This problem is the following:  given a partition of a class $C$ into
disjoint subsets $P_1,\dots,P_k$, determine which piece the unknown
black-box function $c \in C$ belongs to.
Ambainis {\em et al.}~proposed the study of this
problem as an interesting direction for future work in
\cite{AKMPY}.
Note that the problem of computing a single Boolean predicate of an unknown
function $c \in C$ corresponds to having a two-way partition, 
whereas the problem of exact learning corresponds to a partition of $C$
into $|C|$ disjoint pieces.  

We show that for any concept class $C$ and any partition
size $2 \leq k \leq |C|,$ there is a partition of $C$ into $k$
pieces such that the classical and quantum query complexities are polynomially
related.
On the other hand, we also show that for a wide range of partition sizes 
$k$ it is possible for the quantum and classical query complexities 
of learning a $k$-way partition to have a superpolynomial separation.
These results show 
that the structure of the partition plays a more important role 
than the size in determining the relationship between quantum and classical
complexity of learning.

Finally, for the quantum PAC learning model,
we improve the $\Omega({\frac d n})$ lower bound of \cite{RSSG}
on the number of quantum examples which are required to PAC learn
any concept class of Vapnik-Chervonenkis dimension $d$ over $\{0,1\}^n.$
Our new bound of $\Omega(\frac{1}{\epsilon}\log
\frac{1}{\delta}+d +\frac{\sqrt{d}}{\epsilon})$ is not 
far from the known lower bound of Ehrenfeucht {\em et al.}~\cite{EHKV}
of $\Omega(\frac{1}{\epsilon}\log \frac{1}{\delta}+\frac{d}{\epsilon})$ for classical
PAC learning. Since the lower bound of \cite{EHKV} is known to be nearly
optimal for classical PAC learning algorithms (an upper bound of
$O({\frac 1 \eps} \log {\frac 1 \delta} + {\frac d \eps} \log {\frac 1 \eps})$
was given by \cite{BEHW}), our new quantum lower bound is not
far from being the best possible.

\subsection{Organization}
Section~\ref{sec:quantumexact} gives our new quantum algorithm for
exactly learning a black-box function.  
Section~\ref{sec:rel} gives some simple examples
and poses a question about the relation between query complexity
of quantum and classical exact learning.
Section~\ref{sec:partition} gives our
results on the partition learning problem,
and Section~\ref{sec:pac} gives our new lower bound
on the sample complexity of quantum PAC learning.
Section~\ref{sec:conc} concludes with some additional open questions for
further work.

\section{Preliminaries}

\subsection{Learning Preliminaries}

A \emph{concept} $c$ over $\{0,1\}^{n}$ is a Boolean function 
$c: \{0,1\}^{n} \rightarrow \{0,1\}.$ 
Equivalently we may
view a concept as a subset of $\{0,1\}^n$ defined by $\{x \in
\{0,1\}^n : c(x)=1\}.$
A \emph{concept class} ${\cal C} = 
\cup_{n \geq 1} C_n$ is a set of concepts where
$C_n$ consists of those concepts in ${\cal C}$ 
whose domain is $\{0,1\}^n.$
For ease of notation throughout the paper we will omit
the subscript in $C_n$ and simply write $C$ to denote
a collection of concepts over $\{0,1\}^n$.
It will often be useful to think of $C$ as a
$|C| \times 2^n$-binary matrix where rows correspond
to concepts $c \in C,$ columns correspond to 
inputs $x \in \{0,1\}^n$, and the $(i,j)$ entry of the matrix
is the value of the $i$-th concept on the $j$-th input.

We say that a concept class $C$ is {\em 1-sensitive} if it has the property that for 
each input $x$, at least half of all concepts $c \in C$ have $c(x) = 0$ (i.e. each 
column of the matrix $C$ is at most half ones). Given any $C$ it is possible to convert 
it to an equivalent 1-sensitive concept class by flipping the value obtained from any input 
$x$ which has $|\{c : c(x)=1\}| > |\{c : c(x)=0\}|$. This condition on $x$ can simply be checked 
by enumerating all concepts $c$ in $C$ -- without making any queries. In general, we refer 
to the process of flipping the matrix entries which reside in a particular subset of columns 
as performing a {\em column flip}. This notion of 1-sensitivity and a column flip 
was first introduced by \cite{AKMPY}.
 
It is important to note that achieving the effect of 
a column flip in our algorithms involves creating 
and using simulated oracles. In other words, a column flip affects not only the matrix 
corresponding to the set of candidate concepts $C$ but also the result of classical and  
quantum membership queries. Therefore, after a column flip on the subset of inputs $K$, 
a membership query access to the target oracle at one of the inputs in $K$ should be 
considered to be inverted before returned to the algorithm. As remarked in \cite{AKMPY}, in 
both the classical and quantum learning models this can be achieved via some additional 
circuitry which is not significant for our purposes, since we are only interested in 
the query complexity.

\subsection{Classical Learning Models}

The classical model of {\em exact learning from membership queries}
was introduced by Angluin \cite{ANGLUIN} and has since been studied
by several authors \cite{BCG+96,Gavalda94,Hegedus95,HPR+96}.
In this framework, a learning algorithm for $C$ is given query access to a 
black-box oracle $MQ_c$ for the unknown target concept $c \in C$, i.e.
when the learner provides $x \in \{0,1\}^n$ to $MQ_c$ she
receives back the value $c(x).$  A learning algorithm is said to be an 
{\em exact learning algorithm for concept class $C$} if the 
following holds:  for any $c \in C,$
with probability at least $2/3$ the learning algorithm outputs a
Boolean circuit $h$ which is logically equivalent to $c$.  
(Note that a learning algorithm for $C$ ``knows'' the class $C$
but does not know the identity of the target concept $c \in C.$)
The {\em query complexity} of a learning algorithm is 
the number of queries that it makes to $MQ_c$ before outputting $h.$
We will be chiefly concerned in this paper with a quantum version
of the exact learning model, which we describe in 
Section~\ref{sec:quantumexact}. 

In the classical PAC (Probably Approximately Correct) learning model,
which was introduced by Valiant \cite{VALIANT} and subsequently
studied by many authors, the learning algorithm has access to a {\em
random example oracle} $EX(c,\D)$ where $c \in C$ is the unknown target concept
and $\D$ is an unknown probability distribution over $\{0,1\}^n$. 
At each invocation the oracle $EX(c,\D)$ (which takes no inputs) outputs a labeled
example $(x,c(x))$ where $x \in \{0,1\}^n$ is drawn from the distribution $\D.$
An algorithm $A$ is a {\em PAC learning algorithm for concept class $C$} if the
following condition holds:  given any $\eps, \delta > 0,$ for all $c \in C$ and
all distributions $\D$ over $\{0,1\}^n,$ if $A$ is given $\eps,\delta$ and is
given access to $EX(c,\D)$ then with probability at least $1 - \delta$ the
output of $A$ is a Boolean circuit $h: \{0,1\}^n \rightarrow \{0,1\}$ (called a
hypothesis) which satisfies $\Pr_{x \in
\D}[h(x) \neq c(x)] \leq \eps.$
The {\em (classical) sample complexity} of $A$ is the maximum number of 
calls to $EX(c,\D)$ which it makes for any $c \in C$ and any distribution 
$\D.$
In Section~\ref{sec:pac} we will study a quantum version
of the PAC learning model.

\section{\label{sec:quantumexact}Exact Learning with Quantum Membership Queries}

Given any concept $c: \{0,1\}^n \rightarrow \{0,1\},$ 
the \emph{quantum membership oracle} $QMQ_{c}$ is the transformation
which acts on the computational basis states by mapping 
$|x,b\rangle\mapsto|x,b\oplus c(x)\rangle$ where $x \in
\{0,1\}^{n}$ and $b \in \{0,1\}.$
A \emph{quantum exact learning algorithm} for a concept
class $C$ is a sequence of unitary transformations $U_0, QMQ_c,
U_1,QMQ_c,\dots,QMQ_c,U_T$ where each $U_i$ is a fixed unitary
transformation without any dependence on $c.$  
The algorithm must satisfy the following property: for any target
concept $c \in C$ which is used to instantiate the $QMQ$ queries,
a measurement performed on the final state will with
probability at least $2/3$ yield a representation of a
(classical) Boolean circuit $h: \{0,1\}^{n}\rightarrow\{0,1\}$ such that
$h(x)=c(x)$ for all $x\in \{0,1\}^{n}$. 
The {\em quantum query complexity} of the algorithm is $T,$ the number
of invocations of $QMQ_c$.

Note that a
quantum membership oracle $QMQ_{c}$ is identical to the notion of ``a quantum
black-box oracle for $c$'' which has been widely studied 
in e.g.~\cite{BBCMdW,FGGS,GROVER} and many other works.  
Most of this work, however,
focuses on the quantum query complexity of computing a single
bit of information about the unknown oracle, e.g. the OR of all
its output values \cite{GROVER} or the parity of all its output
values \cite{FGGS}.  The quantum exact learning problem 
which we consider in this section 
was proposed in \cite{RSSG} and later
studied in \cite{AKMPY} (where it is called the ``oracle identification
problem'') and in \cite{HMPPR}.

Throughout the paper we write $R(C)$ to denote the minimum query 
complexity
of any classical (randomized) exact learning algorithm for concept class $C.$
We write $Q(C)$ to denote the minimum query complexity of 
any quantum exact learning algorithm for $C.$  We write $N$ to denote 
$2^n,$ the number
of elements in the domain of each $c \in C.$

In Section~\ref{sec:previous} we briefly recap known bounds on the query
complexity of quantum and classical exact learning algorithms. 
In Section~\ref{sec:newquantumalg} we give our new quantum learning
algorithm, prove correctness, and analyze its query complexity.

\subsection{\label{sec:previous}Known bounds on query complexity for exact learning}

We begin by defining a combinatorial parameter $\hat{\gamma}^C$ of a concept
class $C$ which plays an important role in bounds on query complexity
of exact learning algorithms.

\begin{definition}
\label{defgamma}
Let $C$ be a concept class over $\{0,1\}^{n}.$  We define 
\[ \gamma_{a}^{{C}^{\prime}} = \min_{b\in\{0,1\}} |\{c \in {C}^{\prime}:
c(a)=b\}|/ |{C}^{\prime}|,\qquad \mathrm{where}\ a\in \{0,1\}^{n},
{C}^{\prime}\subseteq C \]
    \[ \gamma^{{C}^{\prime}}= \max_{a\in \{0,1\}^{n}}
\gamma_{a}^{{C}^{\prime}},\qquad \mathrm{where}\ {C}^{\prime}\subseteq C\]
    \[ \hat{\gamma}^{C}= \min_{{C}^{\prime}\subseteq C, |{C}^{\prime}|\geq 2} \gamma^{{C}^{\prime}}.\]
\end{definition}
If $C' \subseteq C$ is the set of possible remaining target concepts,
then $\gamma^{C'}$ is the maximum fraction of $C'$ which a (classical)
learning algorithm can be sure of eliminating with a single
query.  Thus, intuitively, the smaller $\hat{\gamma}^C$ is 
the more membership queries should be required to learn $C.$

The following lower and upper bounds on the query complexity
of classical exact learning were established in \cite{BCG+96}:

\begin{theorem} \label{memthm1}
For any concept class $C$ we have
$R(C) = \Omega(\frac{1}{\hat{\gamma}^{C}})$ and $R(C) = \Omega(\log |C|).$ 
\end{theorem}

\begin{theorem} \label{memthm2}
There is a classical exact learning algorithm which
learns any concept class $C$ using 
$O(\frac{\log |C|}{\hat{\gamma}^{C}})$ many queries, so consequently $R(C) = 
O(\frac{\log |C|}{\hat{\gamma}^{C}})$.
\end{theorem}

A quantum analogue of this classical lower bound
was obtained in \cite{RSSG}:

\begin{theorem} \label{memthm3}
For any concept class $C$ over $\{0,1\}^n$ we have $Q(C) = 
\Omega(\frac{1}{\sqrt{\hat{\gamma}^{C}}})$ and $Q(C) = \Omega(\frac{\log |C|}{n})$.
\end{theorem}

Given these results it is natural to seek a quantum analogue
of the classical $O(\frac{\log |C|}{\hat{\gamma}^{C}})$ upper bound.
Hunziker {\em et al.\@} \cite{HMPPR} made the following conjecture:

\begin{conjecture} \label{memconj1}
There is a quantum exact learning algorithm which learns any
concept class $C$ using $O(\frac{\log |C|}{\sqrt{\hat{\gamma}^{C}}})$ quantum membership 
queries.
\end{conjecture}
In Section~\ref{sec:newquantumalg} we prove this conjecture up to a $\log \log |C|$
factor.

Hunziker {\em et al.\@} \cite{HMPPR} also conjectured that there is a quantum
exact learning algorithm which learns any concept class $C$ using 
$O(\sqrt{|C|})$ queries.  This was established by
Ambainis {\em et al.\@} \cite{AKMPY}, who also proved the following result:

\begin{theorem} \label{memthm4} 
There is a quantum exact learning algorithm which learns 
any concept class $C$ with $|C|>N$ using  
$O(\sqrt{N \log |C| \log N}\log \log |C|)$ many queries.
\end{theorem}

\subsection{\label{sec:newquantumalg}A New Quantum Exact Learning Algorithm}

We start with a simple yet useful observation:

\begin{lemma} \label{fact1} For any concept class $C$, there exists an $x \in \{0,1\}^n$ for which at least a 
	$\hat{\gamma}^{C}$ fraction of concepts $c \in C$ satisfy $c(x)=1$. More generally for every subset 
	$C^{\prime}\subseteq C$ with $|C^{\prime}|\geq2$, there exists an input $x$ at which the fraction of 
	concepts in $C'$ yielding $1$ is at least $\gamma^{C^{\prime}}$ (which is at least as large as 
	$\hat{\gamma}^{C}$).
\end{lemma}
\begin{proof} It is sufficient to prove the result in the latter general form. Consider any subset 
	$C^{\prime}\subseteq C$ with $|C^{\prime}|\geq2$. By Definition~\ref{defgamma} we know:
	\begin{itemize}
		\item $\gamma^{C^{\prime}}\geq\hat{\gamma}^{C}$.
		\item At any input $z \in \{0,1\}^n$, the fraction of concepts in $C$ yielding $1$ has to be 
			at least $\gamma_{z}^{C^{\prime}}$.
	\end{itemize}
	Now consider the input $a$ which satisfies $\gamma_{a}^{C^{\prime}}=\gamma^{C^{\prime}}$:
	the fraction of concepts in $C$ yielding $1$ at input $a$ should therefore be at least
	$\gamma^{C^{\prime}}$. Thus taking $x=a$ gives the intended result.	
\end{proof}

The quantity $\hat{\gamma}^C$ can be bounded as follows:
\begin{lemma} \label{lemma1} 
	For any concept class $C$ with $|C| \geq 2$, $\frac{1}{N+1}\leq \gamma^{C}\leq \frac{1}{2}$. 
	This also implies $\frac{1}{N+1}\leq \hat{\gamma}^{C}\leq \frac{1}{2}$ by Definition~\ref{defgamma}.
\end{lemma}
\begin{proof} 
$\gamma^{C}\leq \frac{1}{2}$ is clear from the Definition~\ref{defgamma}.
To prove the other direction we may assume that $\gamma^{C} < \frac{1}{N}$, since otherwise 
the result is obviously true. Therefore $|C|> N$ must hold. Observe that at each input $x$ one
of the following must hold:
\begin{itemize}
	\item The fraction of concepts in $C$ yielding $0$ at $x$ is at least $1-\gamma^{C}$
		and thus the fraction of concepts in $C$ yielding $1$ at $x$ is at most $\gamma^{C}$.
	\item The fraction of concepts in $C$ yielding $1$ at $x$ is at least $1-\gamma^{C}$
		and thus the fraction of concepts in $C$ yielding $0$ at $x$ is at most $\gamma^{C}$.		
\end{itemize}
Hence $C$ can contain at most $\gamma^{C}|C|N$ concepts which are not identically equal to
the concept $c_{\mathsf{maj}}$ defined as follows:
\[ c_{\mathsf{maj}}(x)= \begin{cases}
	0, &\text{if at least half of the concepts in $C$ yield $0$ at $x$;}\\
	1, &\text{otherwise.}
\end{cases}
	\] 
Therefore $C$ must be comprised of these $\gamma^{C}|C|N$ concepts and possibly $c_{\mathsf{maj}}$. 
Thus we obtain: 
\[\gamma^{C}|C|N+1\geq|C| \Longrightarrow \gamma^{C}\geq\frac{|C|-1}{|C|N}\geq\frac{N}{(N+1)N}=\frac{1}{N+1}\]
\end{proof}

\begin{definition} 
A subset of inputs $\mathcal{I} \subseteq \{0,1\}^n$ 
is said to satisfy \emph{the semi-rich 
row condition for $C$} if at least half the concepts in 
$C$ have the property that they
yield $1$ for at least a $\hat{\gamma}^{C}$ fraction of the 
inputs $x$ in $\mathcal{I}$.
\end{definition}

The phrase ``semi-rich row condition'' is used because viewing $C$ as a 
matrix, at least half the rows of $C$ are ``rich'' in 1s (have at least
a $\hat{\gamma}^C$ fraction of 1s) within the columns indexed by inputs
in $\mathcal{I}.$  A simple greedy approach can be used to construct
a set of inputs which satisfies the semi-rich row condition for $C$:

\begin{algorithm}[t]
\caption{Constructing a set of inputs which satisfies the 
semi-rich row condition.}
  \begin{algorithmic}
    \STATE $S\leftarrow\emptyset$, $\mathcal{I}\leftarrow\emptyset$.
    \REPEAT 
    \STATE Perform a column flip on $C \setminus S$ to make $C \setminus S$ be 1-sensitive.
    \STATE $a_{max} \leftarrow$ the input in $\{0,1\}^n \setminus \mathcal{I}$
     at which the highest fraction of concepts in $C\setminus S$ yield $1$.
    \STATE $\mathcal{I} \leftarrow \mathcal{I} \cup \{a_{\max}\}$.
    \STATE $S \leftarrow S \cup$ the set of concepts in the original matrix $C$
    that yield $1$ at input $a_{max}$.
    \UNTIL{$|S|\geq |C|/2$}.
    \STATE Output $\leftarrow\mathcal{I}$.
  \end{algorithmic}
\end{algorithm}

\begin{lemma} \label{lemma2}
Let $C$ be any concept class with $|C| \geq 2.$  Then
Algorithm~1 outputs a set of inputs $\mathcal{I}$ with $|\mathcal{I}| 
\leq {\frac 1 {\hat{\gamma}^C}}$ which satisfies the semi-rich row 
condition for $C$.
\end{lemma}

\begin{proof}
Let $\tau_j |C|$ be the number of concepts in $C \setminus S$ after the
$j$-th execution of the repeat loop in Algorithm~1,
so $\tau_0 = 1.$  Using the first result of Lemma~\ref{fact1}, we obtain  
$\tau_1 \leq 1 - \hat{\gamma}^C$. Now invoking Lemma~\ref{fact1} once again but
this time in its general form, we obtain $\tau_2 \leq (1 - \hat{\gamma}^C)^2$; note that
after the second iteration of the loop, each concept in $S$
will yield 1 on at least half of the elements in $\mathcal{I}.$
Let $j'$ equal $\lfloor {\frac 1 {\hat{\gamma}^C}}\rfloor.$
If Algorithm~1 proceeds for $j'$ iterations through the loop, then it
must be the case that $\tau_{j'} \leq (1 - \hat{\gamma}^C)^{j'}$ and that
each concept in $S$ yields 1 on at least a $\hat{\gamma}^C$ fraction
of the elements in $\mathcal{I}.$  It is easy
to verify that $(1 - x)^{\lfloor 1/x \rfloor} < 1/2$ for $0 < x < {\frac 1 2}.$
We thus have that $\tau_{j'}|C| < |C|/2,$ and consequently $|S| >
|C|/2$, so $\mathcal{I}$ satisfies the semi-rich row condition for
$C$ and the algorithm will terminate before starting
the $(\lfloor {\frac 1 {\hat{\gamma}^C}}\rfloor+1)$-th iteration.

In the case $\hat{\gamma}^{C}\leq \frac{1}{N}$, then the set of all $N$ inputs will 
satisfy the semi-rich row condition for $C$: since any concept which
does not yield $0$ for all inputs actually yields $1$ for at least a 
$\hat{\gamma}^{C}$ fraction of all inputs. Therefore in this case the algorithm
will terminate successfully with an output $|\mathcal{I}| \leq N \leq {\frac 1 {\hat{\gamma}^C}}$. 
Otherwise, since $\hat{\gamma}^C > {\frac 1 N}$, we have that  $j' < N$. This means 
the algorithm never runs out of inputs to add (i.e. $\{0,1\}^N \setminus \mathcal{I}$ is 
nonempty at every iteration).
\end{proof}

Our quantum learning algorithm is given in Algorithm~2.  
Throughout the algorithm the set $S \subseteq C$ should
be viewed as the set of possible target 
concepts that have not yet been eliminated; the algorithm halts when $|S|=1$.
The high-level idea of the algorithm is that in every repetition of the outer loop the size
of $S$ is multiplied by a factor which is at most ${\frac 1 2},$ so at
most $\log|C|$ repetitions of the outer loop are performed.

\begin{algorithm}
  \caption{A Quantum Exact Learning Algorithm.}
  \begin{algorithmic}
    \STATE $S\leftarrow C$.
    \REPEAT
    \STATE Perform a column flip on $S$ to make $S$ $1$-sensitive. Let $\mathcal{K}$ be the set of inputs at which the output is 
    flipped during this procedure.
    \STATE $\mathcal{I} \leftarrow$ The output of Algorithm 1 invoked on the set of concepts $S$.
    \STATE Counter $\leftarrow0$, Success$\leftarrow$\texttt{False}.
    \REPEAT
    \STATE Perform the multi-target subset Grover search on $\mathcal{I}$ using 
    $\frac{9}{2}\lceil\sqrt{|\mathcal{I}|}\rceil$ queries \cite{BBHT}. 
    \STATE $a \leftarrow$ Result of the Grover search.
    \IF{a classical query of the oracle at $a$ yields $1$}
    \STATE $S\leftarrow\{$the concepts in $S$ that yield $1$ at $a\}$, Success$\leftarrow$\texttt{True}.
    \ENDIF
    \STATE Counter$\leftarrow$Counter$+1$.
    \UNTIL{Success \textsc{Or} (Counter$>=\log (3 \log |C|)$)}
    \IF{\textsc{Not} Success}
    \STATE $S \leftarrow$ the set of concepts that yield $0$ for all 
    of the inputs in $\mathcal{I}$. 
    \ENDIF
    \STATE Flip the outputs of concepts in $S$ for all elements in $\mathcal{K}$ to reverse the earlier column flip (thus restoring all concepts in $S$ to their
    original behavior on all inputs).
    \UNTIL{$|S|=1$}.
    \STATE Output $\leftarrow$ A representation of a circuit which
    computes the sole concept $c$ in $S$.
  \end{algorithmic}
\end{algorithm}

\begin{theorem} \label{thm1}
Let $C$ be any concept class with $|C| \geq 2$. Algorithm~2 is a quantum exact learning algorithm for $C$ which
performs $O(\frac{\log |C| \log \log |C|}{\sqrt{\hat{\gamma}^{C}}})$ quantum membership queries.
\end{theorem}
\begin{proof}
Consider a particular iteration of the outer Repeat-Until loop.
The set $S$ is 1-sensitive by virtue of the first step 
(the column flip).
By Lemma~\ref{lemma2}, in the second step of this iteration,
$\mathcal{I}$ becomes a set of at most 
${\frac 1 {\hat{\gamma}^S}}$ many inputs 
which satisfies the semi-rich row condition for $S$.
Consequently, each execution of the Grover search within the inner
Repeat-Until loop uses 
$O(\sqrt{\frac{1}{\hat{\gamma}^{S}}})$ (which is also 
$O(\sqrt{\frac{1}{\hat{\gamma}^{C}}})$) many queries.
Since the inner loop repeats at most $\log (3\log |C|)$ many times,
if we can show that each iteration of the outer loop does indeed 
with high probability (i)
cause the size of $S$ to be multiplied by a factor which is 
at most ${\frac 1 2}$, and (ii) maintain the property
that the target concept is contained in $S$,
then the theorem will be proved.

As shown in \cite{BBHT}, the multi-target Grover search algorithm 
over a space of 
$|\mathcal{I}|$ many inputs using $\frac{9}{2}\lceil\sqrt{|\mathcal{I}|}\rceil$
many queries has the property that if there is any input $a \in \mathcal{I}$
on which the target function yields 1, then the search will output such an $a$
with probability at least ${\frac 1 2}$.
Since the inner loop repeats $\log (3\log |C|)$ many times, we thus have
that if there is any input $a \in \mathcal{I}$ on which
the target concept yields 1, then with probability at least
$1 - {\frac 1 {3 \log |C|}}$ one of the 
$\log (3\log |C|)$ many iterations
of the inner loop will yield such an $a$ and the ``Success''
variable will be set to {\tt True}.  Since the set $S$ is 1-sensitive,
when we eliminate from $S$ all the concepts which
yield 0 at $a$ we will multiply the size of $S$ by at most ${\frac 1 2}$
as desired in this case (and clearly we will not eliminate the
target concept from $S$).  On the other hand, if the set ${\cal I}$
contains no input $a$ on which the target concept yields 1, 
then after $\log (3 \log |C|)$ iterations of the inner loop
we will exit with Success set to {\tt False}, and the concepts
that yield $1$ on any input in ${\cal I}$ will be removed from $S.$  This will
clearly not cause the target to be removed from $S$. Moreover because  
$|{\cal I}| \leq {\frac 1 {\hat{\gamma}^S}}$, any
concept for which even a single input in ${\cal I}$ yields $1$ has
the property that at least a $\hat{\gamma}^S$ fraction of inputs 
in ${\cal I}$ yield $1$. Since ${\cal I}$ satisfies
the semi-rich row condition for $S,$ 
this means that we have eliminated at least half
the concepts in $S$.
Thus, the algorithm will succeed with probability 
at least $(1 - {\frac 1 {3 \log |C|}})^{\log C}$ which is larger than 
$2/3$, and the theorem is proved.
\end{proof}

Recall that $Q(C)$ denotes the optimal query complexity
over all quantum exact learning algorithms for concept class $C$.  
We can show that the query complexity of Algorithm~2 is never much
worse than the optimal query complexity $Q(C)$:
\begin{corollary} \label{cor:comparison} 
For any concept class $C$, Algorithm~2 uses $O(nQ(C)^2 \log \log |C|)$
queries.
\end{corollary}
\begin{proof}
This follows directly from Theorem~\ref{thm1} and the bound $Q(C) 
=\Omega(\frac{\log |C|}{n}+\frac{1}{\sqrt{\hat{\gamma}^{C}}})$
of Theorem~\ref{memthm3}.
\end{proof}

Since $|C| \leq 2^{2^n},$ the bound $O(nQ(C)^2 \log \log |C|)$ is always
$O(n^2 Q(C)^2)$, and thus the query complexity of 
Algorithm~2 is always polynomially related to the query complexity of the
optimal algorithm for any concept class $C.$ 

\subsection{Discussion}

Algorithm~2 can be viewed as a variant of the algorithm of \cite{AKMPY} 
which learns any concept class $C$ from 
$O(\sqrt{N \log |C| \log N}\log \log |C|)$ quantum membership queries.
This algorithm repeatedly performs Grover search over the set of 
all inputs, with the goal each time of 
eliminating at least half of the remaining target concepts.
Instead, our approach is to perform each Grover search only over
sets which satisfy the semi-rich row condition for the remaining set of 
possible target concepts. 
By doing this, we are able to obtain an upper bound on query
complexity in terms of $\hat{\gamma}^{C}$ for every such iteration. 

We observe that our new bound of $O(\frac{\log |C| \log \log |C|}
{\sqrt{\hat{\gamma}^{C}}})$ 
is stronger than the previously obtained upper bound of 
$O(\sqrt{N \log |C| \log N}\log \log |C|)$ from \cite{AKMPY} as
long as ${\frac {\log |C|}{{\hat{\gamma}^{C}}}} = o(N \log N)$. 
Thus, for any concept class $C$ for
which the $O({\frac {\log |C|}{{\hat{\gamma}^{C}}}})$
upper bound of Theorem~\ref{memthm2} on classical
membership query algorithms is nontrivial (i.e.  is less than $N$),
our results give an improvement.

We note that independently Iwama {\em et al.}~\cite{IKRY} have recently 
given a new algorithm for quantum exact learning that uses ideas
similar to the construction of Algorithm~1; however the
analysis is different and their results are incomparable to ours
(their bounds depend only on the number of concepts in $C$ and not
on the combinatorial parameter $\hat{\gamma}^C$).
The main focus of \cite{IKRY} is on obtaining robust learning algorithms
that can learn successfully using noisy oracle queries.

\section{\label{sec:rel}Relations between Query Complexity of Quantum and Classical Exact Learning}

As noted in \cite{RSSG}, combining Theorems~\ref{memthm2} and~\ref{memthm3} 
yields the following:
\begin{corollary} \label{memcor1}
For any concept class $C$, we have 
$Q(C) \leq R(C) = O(nQ(C)^3).$
\end{corollary}

Can tighter bounds relating $R(C)$ and $Q(C)$ be given which hold
for all concept classes $C$?  
While we have not been able to answer this question,
here we make some simple observations and pose a question
which we hope will stimulate further work.

We first observe that the factor $n$ is required in the bound
$R(C) = O(nQ(C)^3)$:
\begin{lemma} \label{lemma4}
For any positive integer $d$ there exists a concept class $C$ 
over $\{0,1\}^n$ with $R(C) = \omega(1)$ 
which has $R(C)  =\Omega(Q(C)^{d})$.
\end{lemma}
\begin{proof}
We assume $d > 1$.  Recall that in the Bernstein-Vazirani problem,
the target concept is an unknown parity function over some
subset of the $n$ Boolean variables $x_1,\dots,x_n$;
Bernstein and Vazirani showed \cite{BERVAZ} that for this
concept class we have $R(C)=n$ whereas $Q(C)=1.$
We thus consider a concept class in which each concept $c$ contains
$n^{1/d}$ copies of the Bernstein-Vazirani problem (each instance
of the problem is over $n^{(d-1)/d}$ variables) as follows:
we view $n$-bit strings $a,x$ as 
\begin{align*}
    &a=(a_{1,1}, a_{1,2},\ldots, a_{1,n^{(d-1)/{d}}}, a_{2,1}, a_{2,2}, \ldots, a_{2,n^{({d-1})/{d}}},\ldots a_{n^{1/d},1}, \ldots, a_{n^{1/d},n^{({d-1})/{d}}})\\
    &x=(x_{1,1}, x_{1,2},\ldots, x_{1,n^{(d-1)/d}}, x_{2,1}, x_{2,2}, \ldots, x_{2,n^{({d-1})/{d}}},\ldots x_{n^{(1/d)},1}, \ldots, x_{n^{1/d},
    n^{(d-1)/{d}}})
  \end{align*}
The class $C$ consists of the set of all $2^n$ concepts: 
\[
f_{a}(x)=\bigvee_{i=1}^{n^{1/d}}
((a_{i,1}, a_{i,2}, \ldots, a_{i,n^{(d-1)/d}})\cdot(x_{i,1}, x_{i,2}, \ldots, x_{i,n^{(d-1)/d}}) \bmod 2)
\]
i.e. $f_a(x)$ equals 1 if any of the $n^{1/d}$ parities
corresponding to the substrings $a_{i,\cdot}$ take value 1 on
the corresponding substring of $x.$

It is easy to see that $n$ queries suffice for a classical algorithm,
and by Theorem~\ref{memthm1} we have $R(C)=\Omega(\log |C|),$ so
$R(C)= \Theta(n).$ 
On the other hand, it is also easy to see that $Q=O(n^{\frac{1}{d}})$
since a quantum algorithm can learn by making $n^{1/d}$ successive
runs of the Bernstein-Vazirani algorithm.

Finally, if $d = 1$ then as shown in \cite{VANDAM} the set $C$ of all
$2^{2^n}$ concepts over $\{0,1\}^n$ has $Q(C)=\Theta(2^n)$
and $R(C) = \Theta(2^n)$.
\end{proof}

The bound $R(C) = O(nQ(C)^3)$ implies that the gap
of Lemma~\ref{lemma4} can only be achieved for concept
classes $C$ which have $R(C)$ small.
However, it is easy to exhibit concept classes which have 
a factor $n$ difference between $R(C)$ and $Q(C)$ 
for a wide range of values of $R(C)$:

\begin{lemma} \label{lemma5}
For any $k$ such that $n - k = \Theta(n),$ there is a concept class
$C$ with $R(C) = \Theta(n2^k)$ and $Q(C) = \Theta(2^k)$.
\end{lemma}
\begin{proof}
The concept class $C$ is defined as follows. 
A concept $c\in C$ corresponds to 
$(a^0,\dots,a^{2^{k}-1})$, where each $a^i$ is a $(n-k)$-bit string.
The concept $c$ maps input $x \in \{0,1\}^n$ to $(a^i \cdot y) \bmod 2$, 
where $i$ is the number between $0$ and $2^k-1$ whose binary
representation is the first $k$ bits of $x$ and $y$ is the $(n-k)$-bit
suffix of $x.$
Since each concept in $C$ is defined uniquely by 
$2^k$ many $(n-k)$-bit strings $a^0,\dots,a^{2^{k}-1}$, there 
are $2^{2^k(n-k)}$ concepts in $C$.
    
Theorem \ref{memthm1} yields $R(C)=\Omega(2^k (n-k))$. 
It is easy to see that in fact $R(C)=\Theta(2^k (n-k))$: 
For each of the $2^k$ parities which one must learn 
(corresponding to the $2^k$ possible prefixes of an input), 
one can learn the $(n-k)$-bit parity with $n-k$ classical queries.

It is also easy to see that by running the Bernstein-Vazirani algorithm 
$2^k$ times (once for each different $k$-bit prefix), 
a quantum algorithm can learn an unknown concept from $C$
exactly using $2^k$ queries, and thus
$Q(C) = O(2^k).$  
The $Q(C) = \Omega({\frac {\log |C|} {n}})$ lower bound of Theorem 
\ref{memthm1} gives us $Q(C) = \Omega({\frac {n - k} n} \cdot 2^k)
= \Omega(2^k),$ and the lemma is proved.
\end{proof}

Based on these observations, we pose the following question:
\begin{question} Does every concept class $C$ satisfy
$R(C) = O(nQ(C)+Q(C)^{2})$?
\end{question}
Note that the example in Lemma \ref{lemma5} 
and the concept class of Grover search \cite{GROVER}: 
$C=\{f_{i}, 0\leq i < N : f_{i}(x)=\delta_{i,x}\}$ saturate this upper bound.

\section{\label{sec:partition}On Learning a Partition of a Concept Class}

\begin{definition} 
Let $C$ be a concept class over $\{0,1\}^n.$ A \emph{partition $\Part$ of $C$} is a collection of nonempty 
disjoint subsets $P_1,\dots,P_k$ whose union yields $C$. 
\end{definition}

In this section we study a different problem,
mentioned by Ambainis {\em et al.}~\cite{AKMPY},
that is more relaxed than exact learning: given a partition
$\Part$ of $C$ and a black-box (quantum or classical) oracle for an unknown
target concept $c$ in $C,$ what is the query complexity of identifying the set $P_i$ in $\Part$ which contains $c$?
It is easy to see that both the exact learning problem (in which $|\Part|=|C|$) and the problem of computing 
some binary property of $c$ (for which $|\Part|=2$) are special cases of this more general problem. 
One can view these problems in the following way: for the exact learning problem the algorithm must obtain all 
$\log |C|$ bits of information about the target concept, whereas for the problem of computing a property of $c$ the 
algorithm must obtain a single bit of information.  In a general instance of the partition problem, the algorithm 
must obtain $\log |\Part|$ bits of information about the target concept.

Given a concept class $C$ and a partition $\Part$ of $C,$ we will write $\RP(C)$ to denote the optimal query
complexity of any classical (randomized) algorithm for the partition problem which outputs the correct answer 
with probability at least $2/3$ for any target concept $c.$  We similarly write $\QP(C)$ to denote the optimal 
complexity of any quantum query algorithm with the same success criterion. 

As noted earlier, for the case $|\Part| = |C|$ we know from Corollary~\ref{memcor1} that the quantities
$\RP(C)$ and $\QP(C)$ are polynomially related for any concept class $C,$ since $\RP(C) = O(n\QP(C)^3).$
On the other extreme, if $|\Part|=2$ then concept classes are known for which $\RP(C)$ and $\QP(C)$ 
are polynomially related (see e.g. \cite{BBCMdW}), and concept classes are also 
known for which there is
an exponential gap \cite{SIMON}. It is thus natural to investigate the relationship between the
size of $|\Part|$ and the existence of a polynomial relationship between $\RP(C)$ and $\QP(C).$

In this section, we show that the number of sets in $|\Part|$ alone (viewed as a function of $|C|$) 
often does not provide sufficient information to determine whether $\RP(C)$ and $\QP(C)$ are polynomially
related. More precisely, in Section~\ref{sec:polyrel} we show that for {\em any} concept class $C$ 
over $\{0,1\}^n$ and any value 
$2 \leq k \leq |C|,$ there is a partition $\Part$ of $C$ with $|\Part|=k$ for which we have $\RP(C)=O(n\QP(C)^{3})$.
On the other hand, in Section~\ref{sec:gap} we show that for a wide range of values of $|\Part|$ (again
as a function of $|C|$), there are concept classes which have a superpolynomial separation 
between $\RP(C)$ and $\QP(C).$ Thus, our results concretely illustrate that the structure of the partition
(rather than the number of the sets in the partition) plays an important role in determining whether the quantum 
and classical query complexities are polynomially related.

\subsection{\label{sec:polyrel}Partition Problems for which Quantum and Classical Complexity are Polynomially Related} 

The following simple lemma extends the cardinality-based lower bounds of 
Theorem~\ref{memthm1} and Theorem~\ref{memthm3} for exact learning to the problem of learning a 
partition:

\begin{lemma} \label{partlemma1}
For any partition $\Part$ of any concept class $C$ over $\{0,1\}^n,$ we have 
$\RP(C)=\Omega(\log |\Part|)$ and $\QP(C)=\Omega(\frac{\log |\Part|}{n})$.
\end{lemma}
\begin{proof} Let $C^{\prime}\subseteq C$  be a concept class formed by taking any single element from 
each subset in the partition 
$\Part$. Learning $\Part$ requires at least as many queries as exact learning the concept class 
$C^{\prime}$, and so the result follows from Theorem~\ref{memthm1} and Theorem~\ref{memthm3}.
\end{proof}

To obtain a partition analogue of the other lower bounds of
Theorems~\ref{memthm1} and~\ref{memthm3}, we define
the following combinatorial parameter which is an analogue of
$\hat{\gamma}^C$:
\begin{definition} \label{defpartgamma}
Let $\mathscr{S}$ be the set of all subsets 
${C}^{\prime}\subseteq C$, $|{C}^{\prime}|\geq 2$ which have the property 
that any subset $C'' \subseteq C'$ with $|C''| \geq {\frac 3 4} |C'|$ must intersect 
at least two subsets in $\Part$. We define $\hat{\gamma}^{C}_{\Part}$ to be
$\hat{\gamma}^{C}_{\Part} :=\min_{{C}^{\prime}\subseteq \mathscr{S}} \gamma^{{C}^{\prime}}$.
\end{definition}
Thus each subset $C'$ in $\mathscr{S}$ has the property that
the partition induced on ${C}^{\prime}$ by $\Part$ contains no subset of size as
large as ${\frac 3 4}|{C}^{\prime}|$. 

The next lemma shows that for each ${C}^{\prime}\in\mathscr{S}$, the lower bounds for exact learning
$R({C}^{\prime}) = \Omega(\frac{1}{{\gamma}^{{C}^{\prime}}})$ and
$Q({C}^{\prime}) = \Omega(\frac{1}{\sqrt{{\gamma}^{{C}^{\prime}}}})$ 
which are implied by Theorems~\ref{memthm1}
and~\ref{memthm3}
extend to the problem of 
learning a partition to yield $\RP({C}^{\prime}) = \Omega(\frac{1}{{\gamma}^{{C}^{\prime}}})$
and $\QP({C}^{\prime}) = \Omega(\frac{1}{\sqrt{{\gamma}^{{C}^{\prime}}}})$. 
By considering the ${C}^{\prime}\in\mathscr{S}$ which minimizes $\gamma^{{C}^{\prime}}$,
we obtain the strongest lower bound (this is the motivation behind Definition~\ref{defpartgamma}).
\begin{lemma} \label{partlemma2}
For any partition $\Part= P_1,\dots,P_k$ of the concept class $C$, we have
$\RP(C)=\Omega(\frac{1}{\hat{\gamma}^{C}_{\Part}})$ and $\QP(C)=\Omega(\frac{1}{\sqrt{\hat{\gamma}^{C}_{\Part}}})$.
\end{lemma}
\begin{proof} 
Let $C' \in \mathscr{S}$ be such that $\hat{\gamma}^{C}_{\Part}=\gamma^{{C}^{\prime}}$.
We consider the problem of learning the partition induced by $\Part$ over $C'$, and shall 
prove the lower bound for this easier problem. 
We may assume without loss of generality that $C^{\prime}$ is 1-sensitive.
    
We first consider the classical case.  We claim that 
there is a partition $\{S_1, S_2\}$ of $C'$ with the property
that each subset $(P_j \cap C')$ is contained entirely in exactly
one of $S_1,S_2$ (i.e. $S_1,S_2$ is a ``coarsening'' of the partition
induced by $\Part$ over $C'$)
which satisfies $\min_{i=1,2}|S_i| > {\frac 1 4}|C'|$.
To see this, we may start with $S_2 = \emptyset$, $S_1 = \cup_{i=1}^k (P_i \cap C') = C'$
and consider a process of ``growing'' $S_2$ by successively removing the smallest piece
$P_j \cap C'$ from $S_1$ and adding it to $S_2.$  W.l.o.g. we may suppose that
$|P_j \cap C'| \leq |P_{j+1} \cap C'|$ for all $j$, so the pieces
$P_{j} \cap C'$ are added to $S_2$ in order of increasing $j = 1,2,\dots.$ Let $t$ be
the index such that adding $P_t \cap C'$ to $S_2$ causes $|S_2|$ to exceed ${\frac 1 4}|C'|$
for the first time.  By Definition~\ref{defpartgamma} it cannot be the case
that adding $P_t \cap C'$ causes $S_2$ to become all of $C'$ (since this would mean
that $P_t \cap C'$ is a subset of size at least ${\frac 3 4}|C'|$ which intersects
only $P_t$); thus it must be the case that after this $t$-th step $S_1$ is still nonempty.
However, it also cannot be the case that after this $t$-th step we have
$|S_1| < {\frac 3 8}|C'|$; for
if this were the case, then after the $t$-th step we would have 
$|P_t \cap C'| > {\frac 3 8}|C'| > |S_1| = \cup_{j=t+1}^k (P_j \cap C')$
and this would violate
the assumption that sets are added to $S_2$ in order of increasing size.

Since $C'$ is 1-sensitive, the ``worst case'' for a learning algorithm is that
each classical query to the target concept (some $c \in C'$) yields 0.
By definition of $\gamma^{C'}$, each such query eliminates at most
$\gamma^{{C}^{\prime}}\cdot|C^{\prime}|$ many possible target concepts from $C'$.
Consequently, after $\lfloor\frac{1/4}{\gamma^{{C}^{\prime}}}\rfloor-1$ 
classical queries, the set of possible target concepts in $C'$ is of size
at least ${\frac 3 4}|C'|,$ and so it must intersect both $S_{1}$ and $S_{2}$. 
It is thus impossible to determine with probability greater than $1/2$ whether
$c$ belongs to $S_1$ or $S_2$, and thus which piece $P_i$ of $\Part$ contains $c$.
This gives the classical lower bound.
    
Our analysis for the quantum case requires some basic definitions and facts
about quantum computing:
    
\begin{definition} If $|\phi\rangle = \sum_z \alpha_z |z \rangle$ and $|\psi\rangle = \sum_z \beta_z |z\rangle$
are two superpositions of basis states, then the {\em Euclidean distance} between $|\phi\rangle$ and $|\psi\rangle$ is
$\||\phi\rangle - |\psi\rangle\| = (\sum_z |\alpha_z - \beta_z|^2)^{1/2}$. The {\em total variation distance} between 
two distributions $\mathcal{D}_1$ and $\mathcal{D}_2$ is defined to be $\sum_x |\mathcal{D}_1(x) - \mathcal{D}_2(x)|$.
\end{definition}

\begin{fact}\label{basfact1} {\bf (See \cite{BERVAZ})}
    Let $|\phi\rangle$ and $|\psi\rangle$ be two unit length superpositions which represent possible states of a
    quantum register. If the Euclidian distance $\||\phi\rangle -|\psi\rangle\|$ is at most $\epsilon$, then performing the same
    observation on $|\phi\rangle$ and $|\psi\rangle$ induces distributions $\mathcal{D}_{\phi}$ and $\mathcal{D}_{\psi}$ which
    have total variation distance at most $4\epsilon$.
\end{fact}
For the quantum lower bound, suppose we have a quantum learning algorithm which
makes at most $T=\lfloor\frac{1/4}{32 \sqrt{\gamma^{{C}^{\prime}}}}\rfloor-1$ quantum membership queries. 
We will use the following result which combines Theorem~6 and Lemma~7 from \cite{RSSG}
(those results are in turn based on
Theorem~6.6 of \cite{BBBV}):

\begin{lemma}[See \cite{RSSG}] \label{memlemma2}
Consider any quantum exact learning algorithm $\mathcal{N}$ which makes $T$
quantum membership queries. Let  $|\phi_T^{c} \rangle$ denote the
state of the quantum register after all $T$ membership queries are performed
in the algorithm, if the target concept is $c$.
Then for any 1-sensitive set $C'$ of concepts
with $|C^{\prime}|\geq 2$ and any $\epsilon > 0$, there is a set 
$S \subseteq C^\prime$ of cardinality at most $T^2 |C^\prime| 
\gamma^{C^{\prime}}/ \epsilon^2$ such that for all 
$c \in C^\prime \setminus S$, we have $\||\phi_T^{\mathbf{0}} \rangle - 
|\phi_T^{c} \rangle\| \leq \epsilon$ (where $\mathbf{0}$ 
denotes the identically $0$ concept).
\end{lemma}
If we take $\epsilon=\frac{1}{32}$, then Lemma \ref{memlemma2} implies that there exists a 
set $S \subseteq C^\prime$ of cardinality less than ${\frac 1 4} \cdot|C^\prime|$ such that 
for all $c \in C^\prime \setminus S$ one has $\||\phi_T^{\mathbf{0}} \rangle - |\phi_T^{c} \rangle\| 
\leq \frac{1}{32}$. 
Consequently by Definition~\ref{defpartgamma} there must
exist two concepts $c_{1}, c_{2} \in C^\prime \setminus S$ with 
$\||\phi_T^{c_{1}} \rangle - |\phi_T^{c_{2}} \rangle\| \leq \frac{1}{16}$
which belong to different subsets $P_i$ and $P_j$
of $\Part$.
By Fact \ref{basfact1}, the probability that our quantum
learning algorithm outputs ``$i$'' can differ by at most
$\frac{1}{4}$ when the target concept is $c_1$ versus $c_2$; but this
contradicts the assumption that the algorithm is correct with probability
at least $2/3$ on all target concepts.  This proves the quantum lower bound.
\end{proof}

Before proving the 
main result of this section, we establish the following result which gives
a sufficient condition for the quantum and classical complexities $\QP(C)$ and $\RP(C)$ 
of learning a partition to be polynomially related.  This result is a generalization
of Corollary~\ref{memcor1}.
\begin{corollary} For a partition $\Part$ over the concept class $C$, if the size of the largest subset
$P_i$ in $\Part$ is less than $\frac{3/4}{\hat{\gamma}^{C}}$, then we have $\RP(C)=O(n\QP(C)^{3})$.
\end{corollary}
\begin{proof} Let 
$C^{\prime}$ be a subset of $C$ for which $\gamma^{C^{\prime}}$ equals $\hat{\gamma}^{C}$. 
We have that $|C^{\prime}|\geq \frac{1}{\hat{\gamma}^{C}}$ by Definition~\ref{defgamma}. Thus any
$\frac{3}{4}$ fraction of $C^{\prime}$ must intersect at least two subsets in $\Part$, so
$C'$ must belong to $\mathscr{S}.$  This forces 
$\hat{\gamma}^{C}_{\Part}=\gamma^{C^{\prime}}=\hat{\gamma}^{C}$.
Moreover, we have that $|\Part|\geq {\frac 4 3}\hat{\gamma}^{C}\cdot |C|$, and we know that  $\frac{1}{N+1}\leq \hat{\gamma}^{C}\leq \frac{1}{2}$ 
by Lemma~\ref{lemma1}. Thus we have $\log |\Part|\geq \log {\frac 4 3} - n +\log |C|,$
and consequently
$\frac{\log |\Part|}{n}> \frac{\log |C|}{n} - 1$. 
Lemmas~\ref{partlemma1} and~\ref{partlemma2} yield
$\QP(C)=\Omega(\frac{\log |\Part|}{n}+\frac{1}{\sqrt{\hat{\gamma}^{C}_{\Part}}})=
\Omega(\frac{\log |C|}{n}+\frac{1}{\sqrt{\hat{\gamma}^{C}}})$.
Combining this with the bound $\RP(C)=O(\frac{\log |C|}{\hat{\gamma}^{C}})$ 
(which clearly follows from Theorem~\ref{memthm2} since 
the partition learning problem is no harder than the exact learning problem), we have that
$\RP(C)$ must be $O(n\QP(C)^{3})$.
\end{proof}

We note here that we could have used any constant $\lambda$ satisfying 
${\frac 2 3} < \lambda < 1$ in Definition~\ref{defpartgamma} in place of 
$3/4,$ and obtained corresponding versions of Lemma~\ref{partlemma2} 
and the above corollary with $\lambda$ in place of $3/4.$

\medskip

Now we prove our main result of this subsection, showing that for {\em any} 
concept class $C$ and any partition size bound $2 \leq k \leq |C|$
there is a partition of $C$ into $k$ pieces such that the 
classical and quantum query complexities are polynomially related:

\begin{theorem} \label{partthm1} 
Let $C$ be any concept class and $k$ any integer satisfying 
$2 \leq k \leq |C|.$  Then there is a partition
$\Part$ of $C$ with $|\Part|=k$ for which we have $\RP(C)=O(n\QP(C)^{3})$.
\end{theorem}

\begin{proof} 
We will show that Algorithm~4 constructs a partition 
$\Part$ with the desired properties.  
Algorithm~4 uses a slightly modified version of Algorithm~1, which we
call Algorithm~3.  Algorithm~3 differs from Algorithm~1 in that if
the input $a_{\max}$ corresponds to a column which is flipped in
the column flip on $C \setminus R,$ then Algorithm~3 augments $R$ by adding
those concepts in the flipped version of $C\setminus R$ which yield
$1$ on $a_{\max}$ (note that by $1$-sensitivity this is fewer
than half of the concepts in $C\setminus R$), whereas
Algorithm~1 adds those concepts which yield $1$ on $a_{\max}$ 
in the unflipped (original) version of $C \setminus R.$
Thus at each stage Algorithm~3
grows the set $R$ by adding at most half of the remaining
concepts in $C\setminus R$; we will need this property later.
The analysis of Algorithm~1 carries over to show that the set
${\cal I}$ of inputs which Algorithm~3 constructs is of size at most
$|{\cal I}| \leq {\frac 1 {\hat{\gamma}^C}}$.

    \begin{algorithm}[t]
    \caption{A slightly modified version of Algorithm 1 to be used in generating a partition.}
    \begin{algorithmic}
        \REQUIRE $C$ is $1$-sensitive.
        \STATE $R\leftarrow\emptyset$, $\mathcal{I}\leftarrow\emptyset$, $\mathcal{J}\leftarrow\emptyset$.
        \REPEAT 
        \STATE Perform a column flip on $C \setminus R$ to make it 1-sensitive;
        call the resulting 1-sensitive matrix $M.$
        \STATE $a_{max} \leftarrow$ the input in $\{0,1\}^n \setminus \mathcal{I}$
         at which the highest fraction of concepts in $C\setminus R$ yield $1$.
        \STATE $\mathcal{I} \leftarrow \mathcal{I} \cup \{a_{\max}\}$.
        \STATE $R \leftarrow R \cup$ the set of concepts in $M$ that yield $1$ at input $a_{max}$.
        \IF{the column corresponding to $a_{\max}$ in $M$ was flipped
        relative to $C$}
        \STATE $\mathcal{J}\leftarrow\mathcal{J}\cup\{a_{\max}\}$.
        \ENDIF
        \UNTIL{$|R|\geq |C|/2$}.
        \STATE Output $\leftarrow(\mathcal{I},\mathcal{J})$.
      \end{algorithmic}
    \end{algorithm}

At each iteration of the outer repeat loop, Algorithm~4 successively
refines the partition $\Q$ until $|\Q|=k$.
Let $C^{\prime} \subseteq C$ be such that $\gamma^{C^{\prime}}=\hat{
\gamma}^{C}$.
The first time Algorithm~4 passes through the inner repeat loop
we will have $|C|=|S|$ and thus Algorithm~3 will be invoked on $C'$.
We will write $C^\circ,C^\star$ to denote these sets $S^\circ,$
$S^\star$ of concepts that are formed out of $C$ in this first iteration.
The final partition $\Part$
will ultimately be a refinement of the partition $\{C^\circ,
C^\star\}$ obtained in this step; we will see later that this will force 
$\hat{\gamma}_{\Part}^C = \hat{\gamma}^C$ (this is why the
first iteration is treated differently than later iterations).

In addition to constructing the partition $\Part,$ 
the execution of Algorithm~4 should also be viewed as a ``memoization''
process in which various sets of inputs ${\cal I}(S),{\cal J}(S)$
and ${\cal K}(S)$ are defined
to correspond to different sets of concepts $S$.
These sets  will be used during the execution of Algorithm~5 later.
Roughly speaking, the division of $S$ in each iteration depends only on the values
on inputs in ${\cal I}(S),$ the set ${\cal J}(S)$ is used to keep track 
of the column flips Algorithm~3 performs, and the set ${\cal K}(S)$ keeps
track of those inputs which need to be flipped to achieve 1-sensitivity.

We now explain the outer loop of Algorithm~4 in more detail.
The algorithm  works in a breadth-first fashion to successively
refine the partition $\Q$, which is initially just $\{C\}$, into
the final partition $\Part$. 
After the first iteration of the outer loop, 
$C$ has been partitioned into $\{C^\circ,C^\star\}$. Similarly, in the
second iteration each of these sets is divided in two to give a four-way partition.
The algorithm continues in this manner until the desired number of elements 
in the partition is reached. 
The main idea of the construction is that each division of a set $S$
(after the first iteration) 
creates two pieces $S^{\circ}$ and $S^{\star}$ of almost equal size as 
we shall describe below.
Because degenerate divisions do not occur, we will see the algorithm
will terminate after at most $O(\log k)$ iterations of the outer 
loop.

Recall from above that each invocation of
Algorithm~3 in Algorithm~4 on a set $S$ of concepts
yields a set ${\cal I}(S)$ of at most ${\frac 1 {\hat{\gamma}^S}}$
many inputs.
By flipping the output of concepts in $S$ at inputs in ${\cal J}(S)$ in
Step 14 of  Algorithm~4, we ensure that the sets 
$S^\circ$ and $S^\star$ defined in steps 15 and 16
correspond precisely
to the sets $C \setminus R$ and $R$ of Algorithm~3 when it terminates.
It thus follows from the termination condition of Algorithm~3
that $|S^{\star}|/ |S| \geq \frac{1}{2}$.
Recall also from the discussion in the first paragraph of this proof
that the last iteration of the
loop of Algorithm~3 adds at most half of the remaining
concepts into the set $R.$
Therefore we have that the set $S^\circ$ in Algorithm~4
must satisfy $|S^{\circ}|/ |S| > \frac{1}{4}$.
It follows from these bounds on $|S^{\circ}|/ |S|$ and $|S^\star|/|S|$
that Algorithm~4
makes at most $O(\log k)$ many iterations through the outer loop.
     
    \begin{algorithm}[t]
        \caption{Constructing a partition for which $\RP(C)$ and $\QP(C)$ are polynomially related.}
        \begin{algorithmic}[1]
            \STATE $\Q\leftarrow\{C\}$
            \REPEAT
            \STATE $\R\leftarrow\emptyset$.
            \REPEAT
            \STATE $S\leftarrow$ an element in $\Q$          
            \IF{$|S|\geq2$}
            \IF{$|C|=|S|$}
            \STATE Let ${\cal K}(S)$ denote the inputs which, if flipped,
			would make $C'$ be 1-sensitive ($C'$ is defined in the 2nd
			paragraph of the proof of Theorem~\ref{partthm1}).  
			Flip the values of concepts
			in $S$ at inputs in ${\cal K}(S).$
            \STATE $(\mathcal{I}(S),\mathcal{J}(S)) \leftarrow$ The output of Algorithm 3 invoked with $C'$.
            \ELSE \STATE Let ${\cal K}(S)$ denote the inputs which, if flipped,
			would make $S$ be 1-sensitive.  Flip the values of concepts
			in $S$ at inputs in ${\cal K}(S).$
            \STATE $(\mathcal{I}(S),\mathcal{J}(S)) \leftarrow$ The output of Algorithm 3 invoked with input $S$.
            \ENDIF
            \STATE Flip the values of concepts in $S$ at those inputs in $\mathcal{J}(S)$.
            \STATE $S^{\circ}\leftarrow \{$the concepts in $S$ that yield $0$ for each $x\in \mathcal{I}(S)\}$.
            \STATE $S^{\star}\leftarrow S\setminus S^{\circ}$.
            \STATE Flip the values of concepts in $S^{\circ}, S^{\star}$ for all elements in $\mathcal{J}(S)$.
            \STATE Flip the values of concepts in $S^{\circ}, S^{\star}$ for all elements in $\mathcal{K}(S)$.
            \STATE $\R\leftarrow \R\cup \{S^{\circ}, S^{\star}\}$. 
            \ELSE
            \STATE $\R\leftarrow \R\cup \{S\}$.
            \ENDIF
            \STATE $\Q\leftarrow \Q\setminus \{S\}$.
            \UNTIL{$\Q=\emptyset$ \textsc{Or} $|\Q|+|\R|=k$.}
            \STATE $\Q\leftarrow \Q\cup \R$.
            \UNTIL{$|\Q|=k$.}
            \STATE $\Part\leftarrow \Q$.
        \end{algorithmic}
    \end{algorithm}
    
It is therefore clear that at each iteration of the main loop of 
Algorithm~4 for which $|S|\geq2$, each
of the sets $S^{\circ}$ and $S^{\star}$ formed from $S$ will be nonempty. 
This ensures that the algorithm will keep producing new elements in the partition until 
$|\Part|=k$ 
is reached. The same argument shows that $C^\circ,C^\star$ are each nonempty and satisfy 
$|C^\circ \cap C'|/|C^{\prime}| > \frac{1}{4}$ and $|C^\star \cap C'|/|C^{\prime}| \geq \frac{1}{2}.$  This implies
that ${C}^{\prime}$ is an element of the set $\mathscr{S}$ of Definition \ref{defpartgamma}: 
any subset $C''\subseteq C'$ with $|C''| \geq {\frac 3 4}|C'|$ 
must intersect both $C^\circ$ and $C^\star,$ and thus must intersect at least two
subsets of $\Part$ since $\Part$ is a refinement of $\{C^\circ,C^\star\}$.
Consequently we have $\hat{\gamma}_{\Part}^{C} = 
\hat{\gamma}^C.$
    
To show that the partition $\Part$ satisfies $\RP(C) = O(n\QP(C)^3),$ 
we now give an analysis of the 
query complexity of learning $\Part$ with both classical and quantum resources.
As we will see, we need to give a classical upper bound and a quantum
lower bound to obtain our goal.
    
In the classical case, we will show that Algorithm~5 makes 
 $O(\frac{\log |\Part|}{\hat{\gamma}^{C}})$ queries and successfully
learns the partition $\Part.$
Using the sets ${\cal I}(S)$ which were defined by
the execution of Algorithm~4, Algorithm~5 makes its way down the
correct branch of the binary tree implicit in the successive refinements
of Algorithm~4 to find the correct
piece of the partition which contains $c$.  More precisely, 
at the end of the $t$-th iteration of the outer loop of Algorithm~5,
the set $S$ which Algorithm~5 has just obtained will be 
identical to the piece $c$ resides in of the
partition constructed by Algorithm~4 at the end of the
$t$-th iteration of its outer loop.
As shown above, it takes $O(\log k) = O(\log |\Part|)$ 
iterations until the subset which
the target concept $c$ lies in is reached.
Moreover, by the same argument in Lemma~\ref{lemma2}, Algorithm~3 always outputs a set of inputs 
$\mathcal{I}(S)$ with size at most ${\frac 1 {\hat{\gamma}^S}} \leq {\frac 1 {\hat{\gamma}^C}}$
when invoked on a set of concepts $S$. Therefore at each of these 
$O(\log |\Part|)$ iterations Algorithm~5 makes at most
$\frac{1}{\hat{\gamma}^{C}}$ many queries.
Thus Algorithm~5 is a classical algorithm that learns $\Part$ using 
 $O(\frac{\log |\Part|}{\hat{\gamma}^{C}})$ queries,
 so we have $\RP(C) = O(\frac{\log |\Part|}{\hat{\gamma}^{C}}).$

    \begin{algorithm}[t]
        \caption{A classical algorithm learning $\Part$.}
        \begin{algorithmic}
            \STATE $S\leftarrow C$
            \REPEAT
            \STATE Flip the values of concepts in $S$ at those inputs in $\mathcal{K}(S)$.
            \STATE Flip the values of concepts in $S$ at those inputs in $\mathcal{J}(S)$.
            \STATE Classically query the given oracle implementing $c$ at all elements in $\mathcal{I}(S)$.
            \STATE $Z\leftarrow$\texttt{True} if $c$ yields $0$ for all elements in $\mathcal{I}(S)$. $Z\leftarrow$\texttt{False} otherwise.
            \IF{$Z$}
            \STATE $S\leftarrow\{$the concepts in $S$ that yield $0$ for each $x\in \mathcal{I}(S)\}$.
            \ELSE
            \STATE $S\leftarrow\{$the concepts in $S$ that yield $1$ for at least one $x\in \mathcal{I}(S)\}$.
            \ENDIF
            \STATE Flip the values of concepts in $S$ at those inputs in $\mathcal{J}(S)$.
            \STATE Flip the values of concepts in $S$ at those inputs in $\mathcal{K}(S)$.
            \UNTIL{$S\in\Part$.}
            \STATE Output$\leftarrow S$.
        \end{algorithmic}
    \end{algorithm}
    
    In the quantum case: since we have
$\hat{\gamma}^{C}_{\Part}=\hat{\gamma}^{C}$, 
by Lemma \ref{partlemma2} any quantum algorithm learning $\Part$ 
should perform $\Omega(\frac{1}{\sqrt{\hat{\gamma}^{C}}})$ quantum membership queries. Combining this 
result with that of Lemma \ref{partlemma1}, we have that $\QP(C)=\Omega(\frac{\log |\Part|}{n}+\frac{1}{\sqrt{\hat{\gamma}^{C}}})$.  Combining this inequality
with the classical upper bound $\RP = O(\frac{\log |\Part|}{\hat{\gamma}^{C}})$ from
Algorithm~5, we have that $\RP(C)=O(n\QP(C)^{3})$ for this partition $\Part$,
and we are done.
\end{proof}

\subsection{\label{sec:gap}A Partition Problem with a Large Quantum-Classical Gap}

In the previous subsection we showed that for any concept class 
and any partition size bound,
there is a partition problem for which the 
classical and quantum query complexities are polynomially related.  
In this section, by adapting a result of Simon \cite{SIMON} we show that 
for a wide range of values of the partition size bound,
it is possible for the classical query complexity to be
superpolynomially larger than the quantum query complexity:

\begin{theorem} 
\label{thm:gap}
Let $n=m + \log m.$
For any value $1\leq \ell < m,$ there is 
a concept class $C$ over $\{0,1\}^{n}$ with
$|C|<2^{m\ell - \ell^2 + \ell + m2^{m-\ell}}$ and a partition $\Part$
of $C$ with $|\mathscr{P}|>2^{m\ell - \ell^2 - \ell}$ such that 
$\RP(C)=\Omega(2^{m-\ell})$ and $\QP(C)=$poly$(m))$.
\end{theorem}

Taking $\ell = m - \alpha(m)$ where $\alpha(m)$ is any function which is
$\omega(\log m)$, we obtain $\RP(C) =m^{\omega(1)}$ whereas 
$\QP(C)=$poly$(m)$, for a superpolynomial separation between
classical and quantum query complexity.
Such a choice of $\ell$ gives $|\mathscr{P}|=2^{\Omega(m)}$ whereas $|C|$ is roughly $2^{m \cdot
2^{\alpha(m)}}$.  Note that the size of $|C|$ can be made to be
$2^{\beta(m)}$ for any slightly superpolynomial function $\beta(m)$
via a suitable choice of $\alpha(m) = \omega(\log m)$. 
This means that viewed as a function of $|C|$, it is possible
for $|\mathscr{P}|$ to be as large as $2^{(\log
|C|)^{\varepsilon(|C|)}}$ for any function
$\varepsilon(\cdot)=o(1)$ and still have the classical query complexity be
superpolynomially larger than the quantum query complexity.

\medskip

\noindent {\bf Proof of Theorem~\ref{thm:gap}:}
We will use a result of Simon \cite{SIMON} who considers
functions $f: \{0,1\}^m \rightarrow \{0,1\}^m.$  
Any such function $f:\{0,1\}^m \rightarrow \{0,1\}^m$ can equivalently
be viewed as a function $\tilde{f}: (\{0,1\}^m \times \{1,\dots,m\})
\rightarrow \{0,1\}$ where $\tilde{f}(x,j)$ equals $f(x)_j$, 
the $j$-th bit of $f(x)$. 
It is easy to see that we can simulate a call to an oracle 
for $f: \{0,1\}^m \rightarrow \{0,1\}^m$ 
by making $m$ membership queries to the oracle for $\tilde{f}$,  in both 
the classical and quantum case. 
This extra factor of $m$ is immaterial for our bounds, so 
we will henceforth discuss functions 
$f$ which map $\{0,1\}^m$ to $\{0,1\}^m.$

We view the input space $\{0,1\}^m$ as the vector space 
$\mathbb{F}^m_2.$
Given a $\ell$-dimensional vector subspace $V \subset \mathbb{F}^m_2,$ 
we say that a function $f:
\{0,1\}^m \rightarrow \{0,1\}^m$ is {\em $V$-invariant}
if the following condition holds:  $f(x) = f(y)$ if and
only if $x = y \oplus v$ for some $v \in V.$  
Thus a $V$-invariant function $f$ is defined by the $2^{m-\ell}$
distinct values it takes on representatives of the 
$2^{m-\ell}$ cosets of $V$.
The concept class $C$ is the class of all functions $f$ which are 
$V$-invariant for some $\ell$-dimensional vector subspace $V$ 
of $\mathbb{F}^{m}_{2}$. 

A simple counting argument shows that there are
\[ N_{m,\ell} = {\frac {(2^m - 1)(2^m - 2)(2^m - 4) \cdots (2^m -
2^{\ell-1})} {(2^\ell - 1)(2^\ell - 2)(2^\ell - 4) \cdots (2^\ell - 2^{\ell-1})}}\]
many $\ell$-dimensional subspaces of $\mathbb{F}^{m}_{2}$. This is because
there are 
$(2^m - 1)(2^m - 2)(2^m - 4) \cdots (2^m - 2^{\ell-1})$
ways to choose an ordered list of $\ell$ linearly independent vectors 
to form a basis for $V$, and given any $V$ there are 
$(2^\ell - 1)(2^\ell - 2)(2^\ell - 4) \cdots (2^\ell - 2^{\ell-1})$ ordered lists
of vectors from $V$ which could serve as a basis.
    
We define the partition $\mathscr{P}$ to divide $C$ into $N_{m,\ell}$ 
equal-size subsets, one for each $\ell$-dimensional vector subspace $V$;
the subset of concepts corresponding to a given $V$ is precisely
those functions which are $V$-invariant.  For any given $\ell$-dimensional
subspace $V$, the number of functions that are $V$-invariant
is
\[
I_{m,\ell} = 2^m(2^{m}-1)(2^m - 2) \cdots (2^{m} - 2^{m-\ell} + 1)
\]
since one can uniquely define such a function by specifying distinct values
to be attained on each of the $2^{m-\ell}$ coset representatives.

Therefore we have $|C|=N_{m,\ell} \cdot I_{m,\ell}$, and it is easy
to check that $2^{m\ell - \ell^2 - \ell} \leq N_{m,\ell} \leq 2^{m\ell - \ell^2 + \ell}$ and 
$I_{m,\ell} \leq
2^{m2^{m-\ell}}$.
It remains only to analyze the quantum and classical query complexities.
    
For the quantum case, it follows easily from Simon's analysis of
his algorithm in \cite{SIMON} that for any $V$-invariant $f,$
each iteration of Simon's algorithm (which requires 
a single quantum query to $f$) will yield a vector
that is independently and uniformly drawn from the $(m-\ell)$-dimensional
subspace $V^{\bot}$.  A standard analysis shows that after $O(m)$
iterations, with very high probability we will have
obtained a set of vectors that span $V^{\bot}$; from these
it is easy to identify $V$ and thus the piece of the
partition to which $f$ belongs.
    
For the classical case, an analysis much like that of
(\cite{SIMON}, Section 3.2)
can be used to show that any classical algorithm which correctly 
identifies the vector subspace $V$ with high  
probability must make $2^{\Omega(m-\ell)}$ many queries; since
the proof is similar to \cite{SIMON} we only sketch the main
ideas here.  We say that a sequence of queries is {\em good}
if it contains two distinct queries which yield the same
output (i.e. two queries $x,y$ which have $(x \oplus y) \in V$),
and otherwise the sequence is {\em bad}.  
The argument of \cite{SIMON} applied to our setting
shows that if the target vector subspace is chosen
uniformly at random, then any classical algorithm
making $M=2^{(m-\ell)/3}$ queries makes a good sequence of queries
with very small probability.
On the other hand, if a sequence of $M=2^{(m-\ell)/3}$ queries is bad, then this
restricts the possibilities for $V$ by establishing 
a set $S$ of ${M \choose 2} < 2^{2(m-\ell)/3}$ many ``forbidden''
vectors from $\mathbb{F}^{m}_{2}$ which must not belong
to the target vector space $V$ (since for each pair of elements $x,y$ 
in $M$ we know that $(x \oplus y) \notin V$).  
Given a fixed nonzero vector $z \in \mathbb{F}^m_{2},$
we have that a random $\ell$-dimensional vector space $W$ contains
$z$ with probability ${\frac {2^\ell - 1}{2^m-1}} < {\frac {2^\ell}{2^m}}$,
and consequently the probability that a random $W$ contains any
element of $S$ is at most $2^{2(m-\ell)/3} \cdot {\frac {2^\ell}{2^m}} = 
2^{(\ell-m)/3}$, which is less than $1/2$ if $\ell<m-6$ (and if 
$\ell\geq m-6$ the bound of the theorem is trivially true).
Thus at least half of the $N_{m,\ell}$ possible 
$\ell$-dimensional vector subspaces are compatible with any given
bad sequence of $2^{(m-\ell)/3}$ queries, so the classical algorithm cannot 
have identified the right subspace with nonnegligible probability.
\qed

\section{\label{sec:pac}Quantum versus Classical PAC Learning}

\subsection{The Quantum PAC Learning Model}

The quantum PAC learning model was introduced by Bshouty and Jackson 
in \cite{BSHJA}.
A quantum PAC learning algorithm is defined analogously to a classical
PAC algorithm, but instead of having
access to a random example oracle $EX(c,\D)$ it can (repeatedly) access a {\em
quantum superposition} of labeled examples.  The goal of constructing a
classical Boolean circuit $h$ which is an $\eps$-approximator for $c$ with
probability $1 - \delta$ is unchanged. More precisely, for $\D$ a distribution
over $\{0,1\}^n$ we say that the {\em quantum example oracle} $QEX(c,\D)$ is a
gate which transforms the computational basis state $|0^{n},0\rangle$ as
follows:

$$|0^{n},0\rangle \mapsto
\sum_{x\in\{0,1\}^{n}} \sqrt{\mathcal{D}(x)}|x,c(x)\rangle.$$ 

We leave the action of a $QEX(c,\D)$ gate undefined on other basis states, and
we require that a quantum PAC learning algorithm may only invoke a $QEX(c,\D)$
oracle on the basis state $|0^n,0\rangle$.  
It is easy to verify (see \cite{BSHJA}) that a $QEX(c,\D)$ oracle
can simulate a classical $EX(c,\D)$ oracle.

As noted in Lemma 6 of \cite{BSHJA},
we may assume without loss of generality (by renumbering qubits) 
that all $QEX(c,\D)$ calls of a 
quantum PAC learning algorithm occur sequentially at the beginning of the 
algorithm's execution and that the $t$-th invocation of $QEX(c,\D)$
affects the qubits $(t-1)(n+1)+1,(t-1)(n+1)+2,\ldots,t(n+1).$ 
After all $T$ $QEX(c,\D)$ queries have been performed, 
the algorithm performs a fixed unitary transformation and then a 
measurement takes place.
(See \cite{BSHJA,RSSG} for more details on the quantum PAC learning model.)
The {\em quantum sample complexity} is
the number of invocations $T$ of $QEX$ which the quantum PAC learning algorithm
performs, 
i.e. the number of $QEX$ gates in the quantum circuit corresponding to
the quantum PAC learning algorithm.

The following definition plays an important role in the sample
complexity of both classical and quantum PAC learning:

\begin{definition} \label{def:VC}
If $C$ is a concept class over some domain $X$ and $W\subseteq X$, we say
that $W$ is \emph{shattered by $C$} if for every $W^{\prime}\subseteq 
W$, there exists a $c\in C$ such that $W^{\prime}=c\cap W$. 
The {\em Vapnik-Chervonenkis dimension of $C$}, $VC-DIM(C)$, is 
the cardinality of the largest $W\subseteq X$ such that $W$ is shattered by $C$.
\end{definition}

\subsection{Known Results on Quantum versus Classical PAC Learning}

The classical sample complexity of PAC learning has been intensively
studied and nearly matching upper and lower bounds are known:
\begin{theorem}
(i) \cite{EHKV} \label{pacthm1}
Any classical $(\epsilon,\delta)$-PAC learning algorithm 
for a non-trivial concept class $C$ of VC dimension $d$ 
must have classical sample complexity 
$\Omega(\frac{1}{\epsilon}\log \frac{1}{\delta}+\frac{d}{\epsilon})$.
(ii) \cite{BEHW} \label{pacthm2}
Any concept class $C$ of VC dimension $d$ can be $(\epsilon, \delta)$-PAC learned
by a classical algorithm with sample complexity $O(\frac{1}{\epsilon}\log
\frac{1}{\delta}+\frac{d}{\epsilon}\log \frac{1}{\epsilon})$.
\end{theorem}

Servedio and Gortler \cite{RSSG} gave a lower bound on the quantum sample
complexity of PAC learning.  They showed that for any concept
class $C$ of VC dimension $d$ over $\{0,1\}^n,$ 
if the distribution $\D$ is uniform over the $d$ examples in some shattered set 
$S$, then even if the learning
algorithm is allowed to make quantum membership queries on any
superposition of inputs in the domain $S$, any 
algorithm which with high probability outputs a high-accuracy hypothesis
with respect to $\D$
must make at least $\Omega({\frac d n})$ many queries. 
Such membership queries can simulate $QEX(c,\D)$ queries since the 
support of $\D$ is $S,$ and thus this gives a lower bound on the sample
complexity of quantum PAC learning with a $QEX$ oracle.

\subsection{Improved Lower Bounds on Quantum Sample Complexity of
PAC Learning}
In this section we give improved lower bounds on the sample
complexity of quantum $(\eps,\delta)$-PAC learning algorithms 
for concept classes $C$ of VC dimension $d.$ These
new bounds nearly match the classical lower bounds
of \cite{EHKV}.

We first note that the $\Omega({\frac d n})$ lower bound 
of \cite{RSSG} can be easily strengthened to $\Omega(d)$: 
\begin{observation} \label{obs:pac}
Let $C$ be any concept class of VC dimension $d$ and let $\D$ be the uniform
distribution over a shattered set $S$ of size $d.$
Then any quantum learning algorithm which (i) can make
quantum membership queries on any superposition of inputs in the domain $S$,
and (ii) with high probability outputs a hypothesis with error
rate at most $\eps = {\frac 1 {10}}$, must make at least ${\frac d {100}}$ 
queries (and consequently the 
sample complexity of PAC learning $C$ with a $QEX$ oracle is $\Omega(d)$).
\end{observation}

Recall that in the exact learning model, the concept class $C$ of all parity 
functions over $n$ Boolean variables has VC dimension $d=n$ yet can be 
exactly learned with one call to a quantum membership oracle
using the Bernstein-Vazirani algorithm \cite{BERVAZ}.
In light of this, we feel that this improvement from $\Omega({\frac d n})$
to $\Omega(d)$ is somewhat unexpected, and may even at first
appear contradictory.  The key to the apparent contradiction is that the 
Bernstein-Vazirani algorithm makes its membership query on a superposition 
of all $2^n$ inputs in $\{0,1\}^n$, not just the $n$ inputs in a 
fixed shattered set $S.$

\medskip

\begin{proof} 
It suffices to slightly sharpen the proof of 
Theorem~4.2 from \cite{RSSG}.  The key observation is that since queries
always have zero amplitude on computational basis states outside
of the shattered set $S,$ the effective value of the
domain size $N$ is $|S|=d$ rather than $|\{0,1\}^n| = 2^n.$
With this modification, at the end of the proof of Theorem~4.2 we obtain
the inequality $N_0 = \sum_{i=0}^{2T} {d \choose i} \geq 2^{d/6}$ where $T$
is the quantum query complexity of the algorithm (instead of the
inequality $\sum_{i=0}^{2T}{2^n \choose i} \geq 2^{d/6}$ which appears in 
\cite{RSSG}).  Now standard tail bounds on binomial coefficients 
(see e.g. Appendix 9 of \cite{KV}) show that $T > {\frac d {100}}.$
\end{proof}

We now give lower bounds on the quantum query complexity of $(\eps,\delta)$-PAC 
learning which depend on $\eps$ and $\delta.$ We require the following
definition and fact:

\begin{definition} A concept class $C$ is said to be \emph{trivial} if either
$C$ contains only one concept, or $C$ contains exactly two concepts 
$c_{0},c_{1}$ with each $x \in \{0,1\}^n$ belonging to exactly one of
$c_0,c_1.$ 
\end{definition}

\begin{fact}[See \cite{SHI}]\label{fact:paq}
Let $|\psi^{(0)}\rangle, |\psi^{(1)}\rangle$ represent states of
a quantum system such that for some measurement $\Pi$ we have 
$\langle\psi^{(0)}|\Pi|\psi^{(0)}\rangle\geq 1-\delta$ and
$\langle\psi^{(1)}|\Pi|\psi^{(1)}\rangle\leq \delta$
for some $\delta > 0.$
Then we have $|\langle\psi^{(0)}|\psi^{(1)}\rangle|\leq 
2 \sqrt{\delta(1-\delta)}$.
\end{fact}

It is clear that a trivial concept class can be learned exactly from
any single (classical) example.  For nontrivial concept classes 
\cite{BEHW} gave a classical 
sample complexity lower bound of $\Omega({\frac 1 \eps} 
\log {\frac 1 \delta})$.  We now extend this bound to the quantum setting:

\begin{lemma}\label{lemmapac}
Any quantum algorithm with a $QEX(c,\D)$ oracle which 
$(\epsilon,\delta)$-learns a non-trivial concept class must have quantum 
sample complexity $\Omega(\frac{1}{\epsilon}\log \frac{1}{\delta})$.
\end{lemma}
\begin{proof} 
Since $C$ is non-trivial, without loss of generality we may assume
that there are two inputs $x^0,x^1$
and two concepts $c_{0}, c_{1}\in C$ such that
$c_{0}(x_{0})=c_{1}(x_{0})=0$ while  $c_{0}(x_{1})=0, c_{1}(x_{1})=1$. 

Let $\D$ be the distribution where 
$\D(x_{0})=1-3\epsilon$ and $\D(x_{1})=3\epsilon$. Under this distribution,
no hypothesis which is $\eps$-accurate for $c_0$ can be $\eps$-accurate
for $c_1$ and vice versa.

Let $|\psi^{(i)}_{T}\rangle$ be the state of the system immediately
after the $T$ queries of $QEX(c_{i},\mathcal{D})$ are performed. Then
we have
\[
\langle\psi^{(i)}_{T} | \underbrace{x_{0},0,x_{0},0,\ldots,x_{0},0}_{
\mathrm{repeated}\ T \mathrm{~times}},0\ldots,0\rangle=(1-3\epsilon)^{T/2},
\quad \mbox{~for~}i=0,1.
\] 
It is easy to see that any other computational basis state 
$|\dots,x_1,b,\dots\rangle$ which has nonzero amplitude in 
$|\psi_T^{(b)}\rangle$ must have zero amplitude in the other possible state
$|\psi_T^{(1-b)}\rangle$, because $c_0$ and $c_1$ disagree on $x_1.$
Consequently we have $\langle\psi^{(0)} |\psi^{(1)}\rangle
=(1-3\epsilon)^{T}$.
If $(1 - 3 \eps)^T > 2 \sqrt{\delta(1-\delta)}$ then 
Fact~\ref{fact:paq} dictates  
there is some output hypothesis which occurs with probability
greater than $\delta$ whether the target is $c_0$ or $c_1$; but this
cannot be the case for an $(\eps,\delta)$-PAC learning algorithm.
Thus we must have $(1-3\epsilon)^{2T}\leq 4\delta$ yielding 
$T=\Omega(\frac{1}{\epsilon}\log \frac{1}{\delta})$.  \end{proof}

Ehrenfeucht {\em et al.\@} \cite{EHKV} obtained a $\Omega({\frac d \eps})$ lower
bound for classical PAC learning by considering a distribution $\D$ which 
distributes $\Theta(\eps)$ weight evenly over all but one of the elements 
in a shattered set. In other words under $\D$ one element in the shattered 
set has weight $1-\Theta(\eps)$ and all the remaining $d-1$ elements has 
equal weight $\frac{\Theta(\eps)}{d-1}$. We use such a distribution to obtain 
the following quantum lower bound (no attempt has been made to optimize constants):

\begin{theorem}\label{thmpac}
Let $C$ be any concept class of VC dimension $d+1$.  Let $\delta = 1/5.$
Then we have that for sufficiently large $d$ (i.e. $d \geq 625$ suffices)
and any $0 < \eps < {\frac 1 {32}},$
any quantum algorithm with a $QEX(c,\D)$ oracle which 
$(\epsilon,\delta)$-learns $C$ must have quantum sample complexity at least 
${\frac {\sqrt{d}} {10000\eps}}$.
\end{theorem}
\begin{proof} 
Let $\{x_{0},x_{1},\ldots,x_{d}\}$ be a set of inputs which is
shattered by $C$. 
We consider the distribution $\mathcal{D}$, first introduced by \cite{EHKV}, 
which has $\mathcal{D}(x_{0})=1-8\epsilon$ and $\mathcal{D}(x_{i})=
\frac{8\epsilon}{d}$ for $i=1,\dots,d.$ 
Let $H(x) = -x \log x - (1-x) \log (1-x)$ denote the binary entropy
function.  
As is noted in \cite{RSSG}, 
there exists a set $s^{1},\ldots,s^{A}$ of $d$-bit strings such that for 
all $i\neq j$ the strings $s^{i}$ and $s^{j}$ differ in at least 
$d/4$ positions where $A \geq 2^{d(1-H(1/4))} > 2^{d/6}$.
For each $i=1,\ldots,A$ let $c_{i}$ be a concept such that (i)
$c_i(x_0) = 0$, and (ii) the 
$d$-bit string $(c_{i}(x_{1}),\ldots,c_{i}(x_{d}))$ is $s^{i}$. 
The existence of such concepts follows from Definition \ref{def:VC}.
Since we have
$\eps < {\frac 1 {32}},$ our quantum PAC learning algorithm 
should successfully distinguish between any two
target concepts $c_{i}$ and $c_{j}$ with confidence at least
$1-\delta= \frac{4}{5}$.
Moreover, without loss of generality we may suppose that 
$\eps < {\frac 1 {100 \sqrt{d}}}$ since otherwise Observation~\ref{obs:pac}
yields the required lower bound.

We shall make use of the following standard inequality:
\begin{equation}
\label{paceqn}
(1-x)^\ell \geq 1-x\ell, \mathrm{if}\ x\ell<1\ \mathrm{for}\ \ell\in
\mathbb{Z}^{+}, x\in\mathbb{R}^{+}.
\end{equation}

Given a target concept $c,$ we write
$|\xi_{i_{1},i_{2},\ldots,i_{t}}\rangle$
to denote the basis state
$$|\xi_{i_{1},i_{2},\ldots,i_{t}}\rangle = 
|x_{i_{1}},c(x_{i_{1}}),x_{i_{2}},c(x_{i_{2}}),\ldots x_{i_{t}},c(x_{i_{t}})
\rangle.$$
We define the state
$|\phi_{t}\rangle$ to be
$$
|\phi_t\rangle = (1-8\epsilon)^{t/2}|\xi_{0,0,\ldots,0}\rangle+
(1-8\epsilon)^{\frac{t-1}{2}} \sqrt{\frac{8\epsilon}{d}}\sum_{i=1}^{d}(|\xi_{i,0,0,\ldots,0}\rangle+|\xi_{0,i,0,\ldots,0}\rangle+\ldots+|\xi_{0,0,\ldots,i}\rangle) + \alpha |z\rangle.
$$
Here $|z\rangle$ is some canonical basis state which is distinct
from, and hence orthogonal to, all states of the form $|\xi_{i_1,i_2,\dots,i_t}\rangle$,
e.g. we could take $z = |x_1, c(x_1), x_1, 1-c(x_1),0,0,\ldots,0 \rangle$.
The scalar $\alpha$ is a suitable normalizing coefficient so that the Euclidean
norm of $|\phi_{t}\rangle$ is 1.

Let $|\psi_{t}\rangle$ denote the state of the quantum register
after $t$ invocations of $QEX(c,\mathcal{D})$ have occured. 
It is easy to see from the definition of the $QEX(c,\D)$ oracle
that the amplitude of $|\psi_t\rangle$ on the basis state 
$|\xi_{0,\dots,0}\rangle$ will be $( 1 - 8 \eps)^{t/2}$,
and that for each of the $td$ many basis states
$|\xi_{0,0,\dots,0,i,0,\dots,0}\rangle$ (where $i$ ranges over all
$t$ positions and ranges in value from $1$ to $d$)
the amplitude of $|\psi_t\rangle$ on this basis state will be 
$(1 - 8\eps)^{{\frac {t-1} 2}} \sqrt{\frac {8 \eps} d}.$
We thus have that
$\langle\psi_{t}|\phi_{t}\rangle=
(1-8\epsilon)^{t}\left(1+\frac{8t\epsilon}{1-8\epsilon}\right).$

If we let $t = {\frac 1 {100 \eps \sqrt{d}}}$ (note that
$t \geq 1$ by our assumption that $\eps < {\frac 1 {100 \sqrt{d}}}$),
we then have that $(1-8\epsilon)^{t} > (1-\frac{1}{12\sqrt{d}})$
by (\ref{paceqn}), and it is easy to check
that $(1+\frac{8t\epsilon}{1-8\epsilon})>1+\frac{1}{12\sqrt{d}}$
from the bounds on $\eps$ and $d$ in the theorem statement.
We thus have that
\begin{eqnarray}
\langle\psi_{t}|\phi_{t}\rangle > 1 - {\frac 1 {144d}}.
\label{eq:pacclose}
\end{eqnarray}

Now let us consider what happens if we replace each successive
block of $t = {\frac 1 {100 \eps \sqrt{d}}}$ invocations of the $QEX(c,\D)$
oracle in our PAC learning algorithm with the transformation
\[Q: |{0}^{t(n+1)}\rangle \mapsto |\phi_{t}\rangle.\]
If the learning algorithm makes a total of $T$ calls to $QEX(c,\D)$ then 
we perform $T/t$ replacements.
After all $T/t$ calls to $Q$ in the modified algorithm, the 
initial state $|0\ldots0\rangle$ evolves 
into the following state $|\varphi\rangle$:
$$|\varphi\rangle = 
\underbrace{|\phi_{t}\rangle\ldots|\phi_{t}\rangle}_{\mbox{$T/t$ times}}|0
\ldots0\rangle.$$

By Equation~(\ref{eq:pacclose}), we have that 
$\langle \phi_T | \varphi \rangle > (1 - {\frac 1 {144d}})^{T/t}.$
If $T/t \leq d/100$ (i.e. if $T \leq {\frac {\sqrt{d}} {10000\eps}}$), 
then by~(\ref{paceqn}) this lower bound is at least ${\frac {143}{144}},$
and this implies (since the original algorithm with $T$ many
$QEX(c,\D)$ calls was successful on each target $c^i$
with probability at least $4/5$)
that the modified algorithm which makes at most $d/100$ many 
calls to $Q$ is successful with probability
at least $2/3$.
However, the exact same polynomial-based argument which underlies
the $\Omega(d/n)$ lower bound for PAC learning proved in \cite{RSSG} (and the
improved ${\frac d {100}}$ lower bound of Observation~\ref{obs:pac})
implies that it is impossible for our modified algorithm, which makes
at most $d/100$ many calls to $Q$, to succeed on each target
$c_i$ with probability at least $2/3$.
(The crux of that proof is that each invocation of a black-box oracle 
for $c$ increases the degree of the polynomial associated to 
the coefficient of each basis state by at most one.  This
property is easily seen to hold for $Q$ as well -- after $r$
queries to the $Q$ oracle, the coefficient of each basis state
can be expressed as a degree-$r$ polynomial in the 
indeterminates $c(x_1),\dots,c(x_d)$.)
This proves that we must have $T/t > d/100,$ which gives the
conclusion of the theorem.
\end{proof}

Combining our results, we obtain the following quantum version of 
the classical $\Omega({\frac 1 \eps}\log {\frac 1 \delta} + 
{\frac d \eps})$ bound:
\begin{theorem}\label{thmpac1}
Any quantum $(\epsilon,\delta)$-PAC learning algorithm for a concept
class of VC dimension $d$ must make at least
$\Omega(\frac{1}{\epsilon}\log \frac{1}{\delta}+d+\frac{\sqrt{d}}{\epsilon})$
calls to the $QEX$ oracle.
\end{theorem}

\section{\label{sec:conc}Future Work}
Several natural questions for future work suggest themselves.  
For the quantum exact learning model, is it possible to
get rid of the $\log \log |C|$ factor in our algorithm's upper
bound and thus prove the conjecture of Hunziker {\em et al.}~\cite{HMPPR} exactly?
For the partitions problem, can we extend the range of partition sizes
(as a function of $|C|$) for which there can be a superpolynomial
separation between the quantum and classical query complexity of
learning the partition?
Finally, for the PAC learning model, a natural goal is to
strengthen our $\Omega({\frac {\sqrt{d}} \eps})$ lower bound
on sample complexity to $\Omega({\frac d \eps})$ and thus
match the lower bound of \cite{EHKV} for classical PAC learning.

\end{document}